\documentclass[final,3p,twocolumn]{elsarticle}

\usepackage{amssymb}
\usepackage{amsthm,amsmath}
\usepackage{caption}
\usepackage{lmodern}
\usepackage{lineno}
\journal{arXiv}

\begin{document}
\begin{frontmatter}


\title{475\textdegree C aging embrittlement of partially recrystallized FeCrAl ODS ferritic steels after simulated tube process}


\author[1]{Zhexian Zhang\corref{cor1}\fnref{fn1}}
\ead{zzhan124@utk.edu}
\author[2]{Daniel Morrall\fnref{fn2}}
\author[1]{Kiyohiro Yabuuchi}

\affiliation[1]{organization={Institute of Advanced Energy, Kyoto University},
            addressline={Gokasho}, 
            city={Uji},
            postcode={611-0011}, 
            state={Kyoto},
            country={Japan}}
            
 \affiliation[2]{organization={Graduate School of Energy Science, Kyoto University},
             addressline={Yoshida-honmachi, Sakyo-ku},
             city={Kyoto},
             postcode={606-8501},
             state={Kyoto},
             country={Japan}}

\cortext[cor1]{corresponding author}
\fntext[fn1,fn2]{This work was done by Zhexian Zhang, and Daniel Morrall when they were staff of Institute of Advanced Energy, Kyoto University and student of  Graduate School of Energy Science, Kyoto University, respectively. Zhexian Zhang is now a visiting researcher in the University of Tennessee Knoxville.}

\begin{abstract}
Tube processing and aging effects in FeCrAl ODS steels are investigated in four mechanical alloyed ferritic ODS steels, Fe15Cr (SP2), Fe15Cr5Al (SP4), Fe15Cr7Al (SP7) and Fe18Cr7Al (SP11). These steels were made into 0.3mm thick plates by simulated tube processing (STP). Strengthening after partial recrystallization was achieved after the last cold rolling and heat treatment step. However, the ductility reduced about one third of the as-extruded steels. The STPed steels were aged at 475\textdegree C in sealed vacuum tubes up to 2000 hrs and 10000 hrs, respectively. The yield stress and elongation were investigated by tensile tests. The results revealed that all the STPed steels fractured in a ductile manner irrespective of aging conditions. Aging hardening and ductility reduction in STPed steels are similar to as-extruded ones. The STPed ODS steels showed similar ageing embrittlement resistance as as-extruded steels, but much higher than the non-ODS steels. The aging hardening based on cut-through and bow-pass mechanisms were discussed. The time dependent hardening of overaged steel ($\beta^\prime$ only) was analyzed as well.
\end{abstract}

%


\end{frontmatter}


\section{Introduction} \label{sec:1}
FeCrAl ferritic alloys are considered as promising candidate materials for accident tolerant fuel (ATF) cladding in the designs for suppressing hydrogen generation reactions with hot water in light water reactors (LWR) at extreme high temperatures\citep{ridgeAccidentTolerantFuel2014,pasamehmetogluStateoftheArtReportLight2018,zinkleAccidentTolerantFuels2014}. However, the neutron penalty caused by larger absorption section of Fe atoms, which will reduce the neutron efficiency in comparison to conventional zircaloy system, has to be relieved by reducing the FeCrAl cladding wall thickness and/or enhancing the uranium enrichment of the nuclear fuel. For example, the thickness of iron-based alloy claddings was calculated to be limited to ~300 $ \mu m $ with 4.78\% U235 fuel enrichment to match the same cycle length of Zircaloy without changing other fuel pin geometries\citep{georgeFullcoreAnalysisFeCrAl2019}. In general, reducing the cladding tube thickness is preferred than enhancing uranium enrichment upon the technical feasibility and economy preference\citep{terraniAdvancedOxidationresistantIronbased2014}. 

Nevertheless, the structural integrity of claddings requires a minimum thickness to meet the strength demand for safety consideration. To compensate for the strength loss by the thickness reduction of cladding wall and to enhance the strength at elevated temperatures, the strategy of oxide dispersion strengthening (ODS) was adopted in the development of FeCrAl ATF cladding materials. Based on this strategy, several Japan national programs of FeCrAl ODS ferritic steels R\& D were conducted\citep{yamashitaTechnicalBasisAccident2017,sakamotoDevelopmentAccidentTolerant2021}. These programs are on the bases of the knowledges accumulated by the Japanese leading programs of R\& D of ODS ferritic\/ martensitic steels with\citep{kimuraDevelopmentFuelClad2003}\citep{kimuraHighBurnupFuel2007,kimuraSuperODSSteels2009} and without\citep{ukaiPerspectiveODSAlloys2002} Al-addition for the applications to core components of Gen IV fast reactors. After decades of investigations, the FeCrAl ODS ferritic steels developed in Japan have demonstrated excellent properties, particularly the creep resistance\citep{ohtsukaImprovement9CrODSMartensitic2004,kimuraDevelopmentAddedHighCr2011,yanoUltrahighTemperatureTensile2017}, oxidation/corrosion resistances\citep{choCorrosionResistanceHighCr2007,isselinCorrosionBehaviour162010,leeInfluenceAlloyComposition2011,liuEffectsCrConcentrations2013,jeStressCorrosionCracking2014}
  and radiation tolerance\citep{kasadaSuperiorRadiationResistance2010,songHeliumBubbleFormation2017,songIonirradiationHardeningAccompanied2018,haEffectRecrystallizationIonirradiation2015}. 

However, the increased ultra-high strength of FeCrAl ODS steels also brought about difficulty in the fabrication processes such as cold rolling and pilger milling\citep{maloyViabilityThinWall2016,aghamiriMicrostructureTextureEvolution2020}. This concern was proposed to be overcome by recrystallization treatment, which reduces the yield strength of ODS steels during tubing\citep{ukaiTubeManufacturingCharacterization2000}. According to the previous study on FeCr(Al) ODS ferritic steels\citep{haRecrystallizationBehaviorOxide2014}, the typical recrystallization temperatures after hot extrusion ranges from 1050\textdegree C to 1400\textdegree C, depending on the alloy compositions and pre-mechanical processing (forge, rolling, etc). The on-set recrystallization temperature of FeCrAl ODS ferritic steels could be significantly reduced by the degree of cold work\citep{ukaiTubeManufacturingCharacterization2000}. The reduction of yield stress could reach to 25\% \- 50\%  after fully recrystallization, while the property of total tensile elongation would barely change\citep{haEffectColdRolling2019}. 

In the tubing process, a four-cycle cold-rolling (CR) and heat-treatment (HT) process was first designed in the R\& D of Fe9Cr ODS alloy\citep{ukaiDevelopment9CrODSMartensitic2002,ukaiNanomesoscopicStructureControl2007}. Later, this processing route was applied to general tubing of FeCrAl ODS steels\citep{sakamotoDevelopmentAccidentTolerant2021}\citep{yanoEffectsThermalAging2021}. The thickness reduction was about 50\% in each CR step. However, the temperatures of the HT vary in intermediate and last steps. The intermediate HT temperatures were designed to be a little below the recrystallization temperature to induce only recovery to avoid cracking in further rolling\citep{naritaCharacterizationRecrystallization12Cr2013}. The recrystallization was only designed in the last step to produce equiaxed grains (however, most of the grains still elongated to the rolling direction, but generally the anisotropy was greatly reduced) and reduce the overall yielding strength. Although recrystallization can effectively reduce the yielding strength, it has been found that repeating recrystallization will be retard by the previous recrystallization\citep{naritaCharacterizationRecrystallization12Cr2013,naritaDevelopmentTwostepSoftening2004,lengEffectsTwostepCold2012}. This is the reason that recrystallization was only designed in the last step in tubing processing.

During the service of FeCrAl tube claddings in reactors, the steels may suffer aging embrittlement under the operation temperatures. This aging effect is owing to the formation of various fine precipitates which cause the deterioration of mechanical properties. According to the recent publications based on atom probe tomography (APT), age-hardening could be brought about by both Cr-rich $\alpha^\prime$ phases\citep{capdevilaPhaseSeparationPM2008,novyAtomicScaleAnalysis2009,capdevilaInfluenceRecrystallizationPhase2012,briggsCombinedAPTSANS2017,edmondsonIrradiationenhancedPrecipitationModel2016,ejenstamMicrostructuralStabilityFe2015} and\/or (Al, Ti)-rich $\beta^\prime$ phases\citep{capdevilaStrengtheningIntermetallicNanoprecipitation2016,sangEarlystageThermalAgeing2020,douEffectsContentsTi2020,douAgehardeningMechanisms15Cr2020} depending on the concentration of Cr and Al in the FeCrAl ODS steels. As for the effect of Al addition on age-hardening, Kobayashi and Takasugi\citep{kobayashiMapping475Embrittlement2010} investigated the age-hardening of diffused FeCrAl alloys with multiple concentrations and showed that the addition of Al shifted the Fe-Cr miscibility boundary to a higher Cr concentration, consequently hinders the $\alpha$-$\alpha^\prime$ phase separation. Particularly in the study by Dou et al, when Al is $>$7wt\%, no $\alpha^\prime$ phase was formed in Fe15Cr-(7,9)Al ODS steels\citep{douEffectsContentsTi2020}, while the Al additions will enhance the formation of Ti-Al enriched $\beta^\prime$ precipitates causing age-hardening without the occurrence of $\alpha$-$\alpha^\prime$ phase separation\citep{capdevilaStrengtheningIntermetallicNanoprecipitation2016}. Sang et al\citep{sangEarlystageThermalAgeing2020} reported that the $\beta^\prime$ could form in early stage during thermal aging in high Al ODS steels. The age-hardening of as-extruded FeCrAl ODS steels has been investigated at 475\textdegree C up to 9000 hrs\citep{hanEffectCrContents2016}, while the aged FeCrAl ODS steels after tubing process have been investigated by Yano et al\citep{yanoEffectsThermalAging2021}. The aging of high-Al FeCrAl ODS steels were also investigated by Maji et al\citep{majiMicrostructuralStabilityIntermetallic2021} where again no $\alpha^\prime$ but Al enriched precipitates (mainly FeAl and Fe\textsubscript{3}Al) were formed in these steels. 

The solution to the aging embrittlement is to optimize the composition of FeCrAl steels. The concentrations of Cr and Al are required to satisfy both the oxidation resistance at extreme high temperature, and low aging rate at reactor operation temperatures. To this end, the “SP” series of FeCrAl ODS steels were designed and produced, with the Cr between 12 and 18wt\%, and the Al of 0, 5~9wt\%. In this range of Cr concentration, $\alpha^\prime$ will form through a special nucleation and growth way (where the Cr concentration in precipitates and size increase simultaneously)\citep{capdevilaPhaseSeparationPM2008}, while Al will hinder this process\citep{kimuraTwofoldAgehardeningMechanism2023}. These researches focusing on the SP-series FeCrAl steels have been intensively carried out and summarized in a recent review\citep{ukaiAlloyDesignCharacterization2023}.

This study belongs to the bunch of the research work on the “SP” series FeCrAl ODS steels. The objective is to investigate the aging behavior of the selected FeCrAl steels after simulated tube fabrication processing (STP). Different to conventional design with recrystallization only in the last HT of tubing process, we tentatively designed the full recrystallization in the 3rd HT and partially recrystallization in the last HT of the STP. This design aims to reduce the rolling difficulty in the 4th CR and increase the final strength of the steels than conventional fully recrystallized FeCrAl tube. The STPed steels were thermal aged at 475\textdegree C up to 2000 hrs and 10000 hrs. The mechanical properties were investigated by means of uniaxial tensile test. A comparison of the aging hardening behavior was made between the as-extruded and STPed FeCrAl ODS steels. The hardening rates were analyzed by combination of a precipitation kinetics model and two-fold hardening mechanism. The (Cr, Al) concentration dependence on hardening was discussed as well.

\begin{figure}[t] 
\centering
\includegraphics[width=0.9\linewidth]{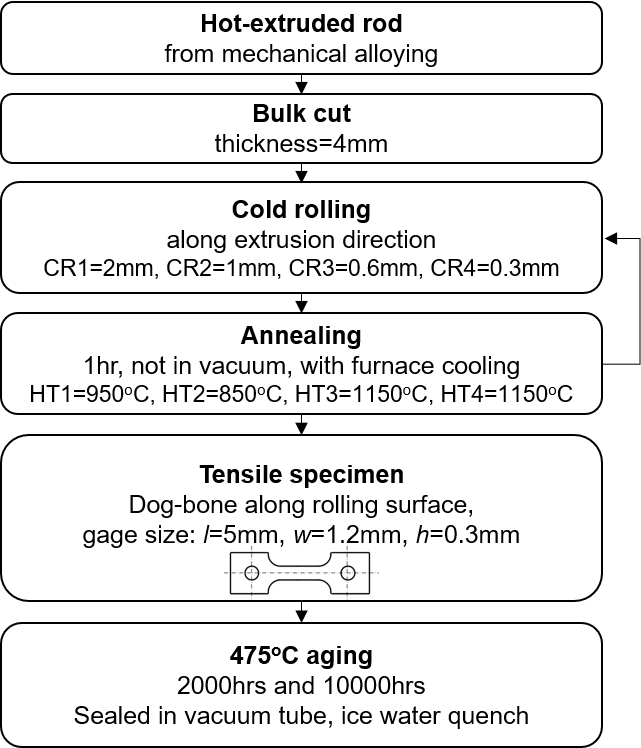}
\caption{The workflow of experimental procedure includes simulated tube processing and thermal ageing together with each measurement and observation. }\label{fig:1}
\end{figure}

\begin{figure*}[btp] 
\centering
\includegraphics[width=\textwidth]{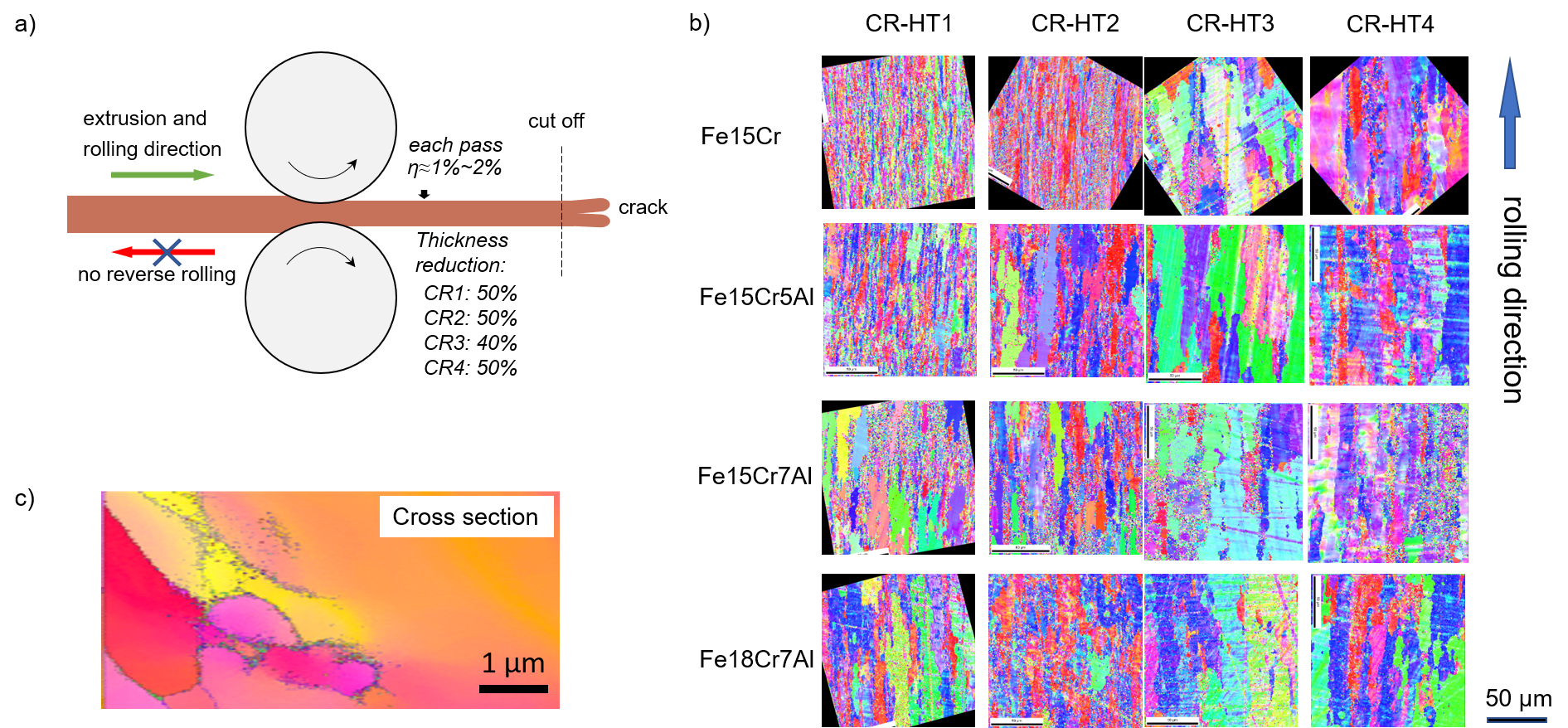}
\caption{a) Schematic view of the cold rolling in the simulated tube processing, b) IPF of the grain morphologies of four FeCrAl ODS ferritic steels rolling surface after each cold rolling and heat treatment (CR-HT) cycle, and c) the transmitted EBSD (TKD) image of cross-section of the fine subgrains of Fe15Cr after ageing for 2000 hr. The upward of conventional EBSD IPF is parallel to the rolling direction. The TKD specimen is normal to the rolling direction.}\label{fig:2}
\end{figure*}

\begin{figure*}[btp] 
\centering
\includegraphics[width=0.9\textwidth]{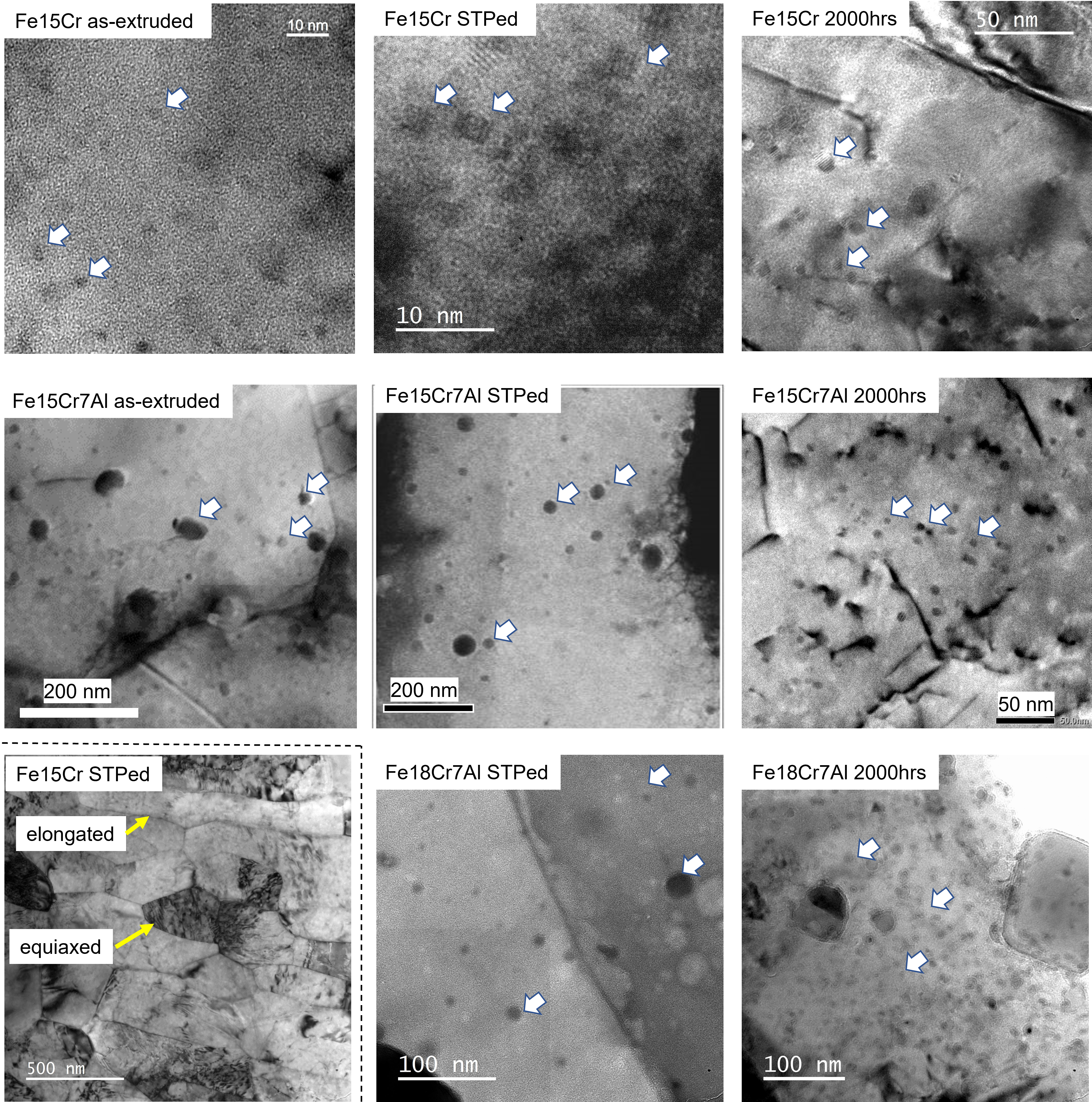}
\caption{The dispersion morphology of nano-particles in FeCrAl ODS ferritic steels at the conditions of as-extruded, simulated tubing processing (STP) and 475\textdegree C 2000hrs aging. The wight arrows indicate oxide particles. The grain morphology of STPed Fe15Cr was shown as well. There are both elongated and equiaxed grains, indicating partially recrystallization in this material.}\label{fig:3}
\end{figure*}

\begin{figure}[bt] 
\centering
\includegraphics[width=\linewidth]{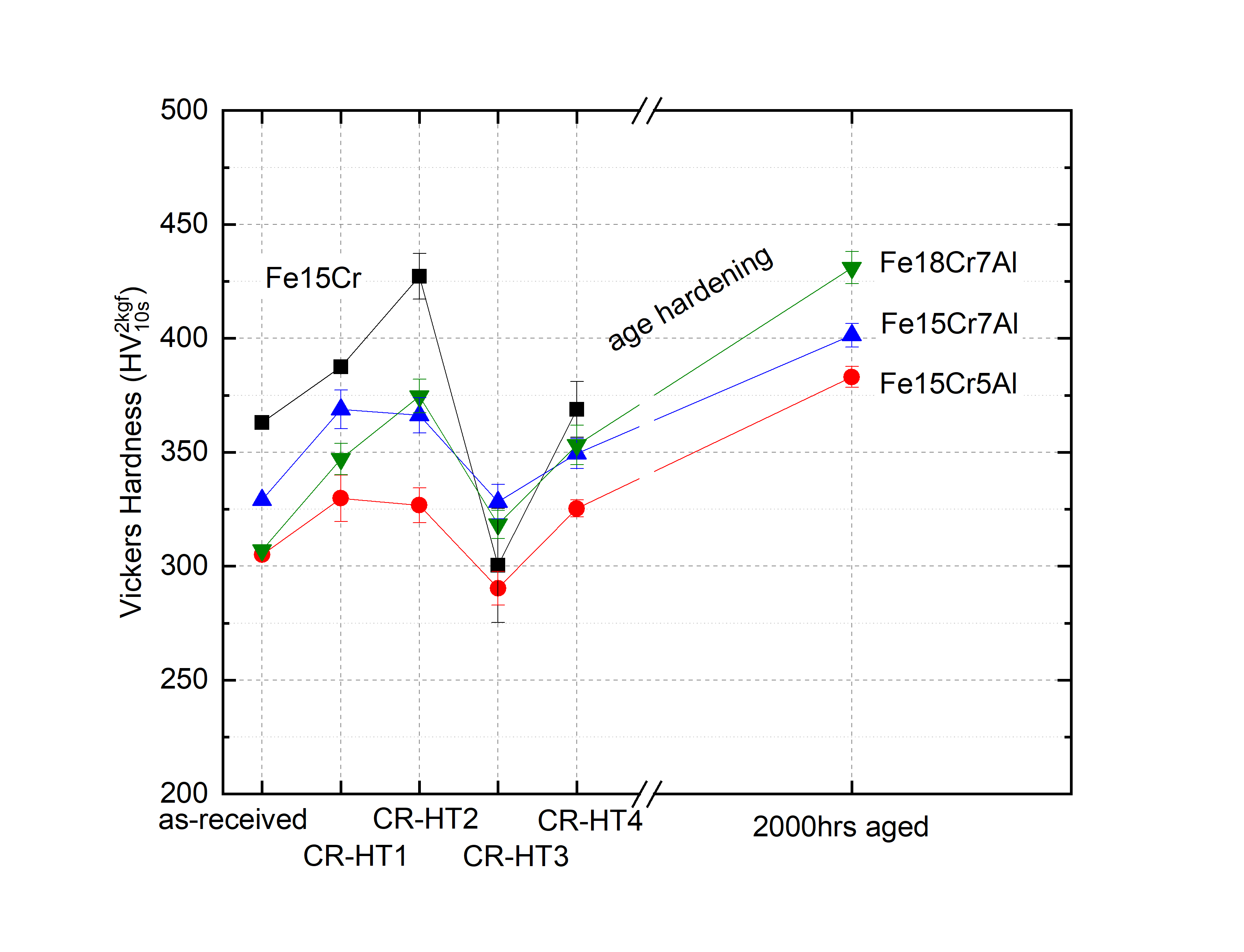}
\caption{The Vickers hardness of FeCrAl ODS ferritic steels during simulated tubing processing after each cold rolling and heat treatment (CR-HT) cycle. All the hardness were measured on the rolling surface.}\label{fig:4}
\end{figure}

\begin{figure*}[btp] 
\centering
\includegraphics[width=0.8\textwidth]{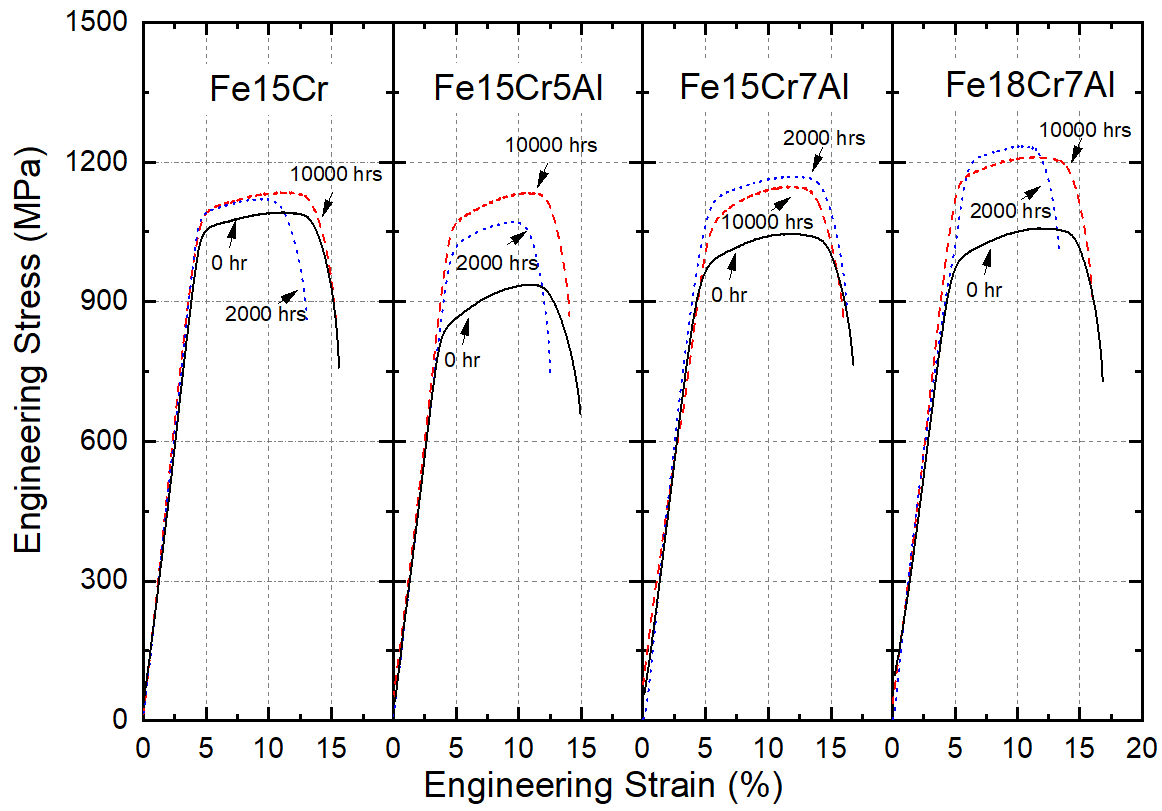}
\caption{The engineering stress-strain curves of STPed Fe15Cr, Fe15Cr5Al, Fe15Cr7Al, Fe18Cr7Al ODS steels before and after ageing at 475\textdegree C to 2000 hrs and 10000 hrs. Note that the elastic component includes the elongation of machine assembly.}\label{fig:5}
\end{figure*}

\begin{figure*}[btp] 
\centering
\includegraphics[width=0.9\textwidth]{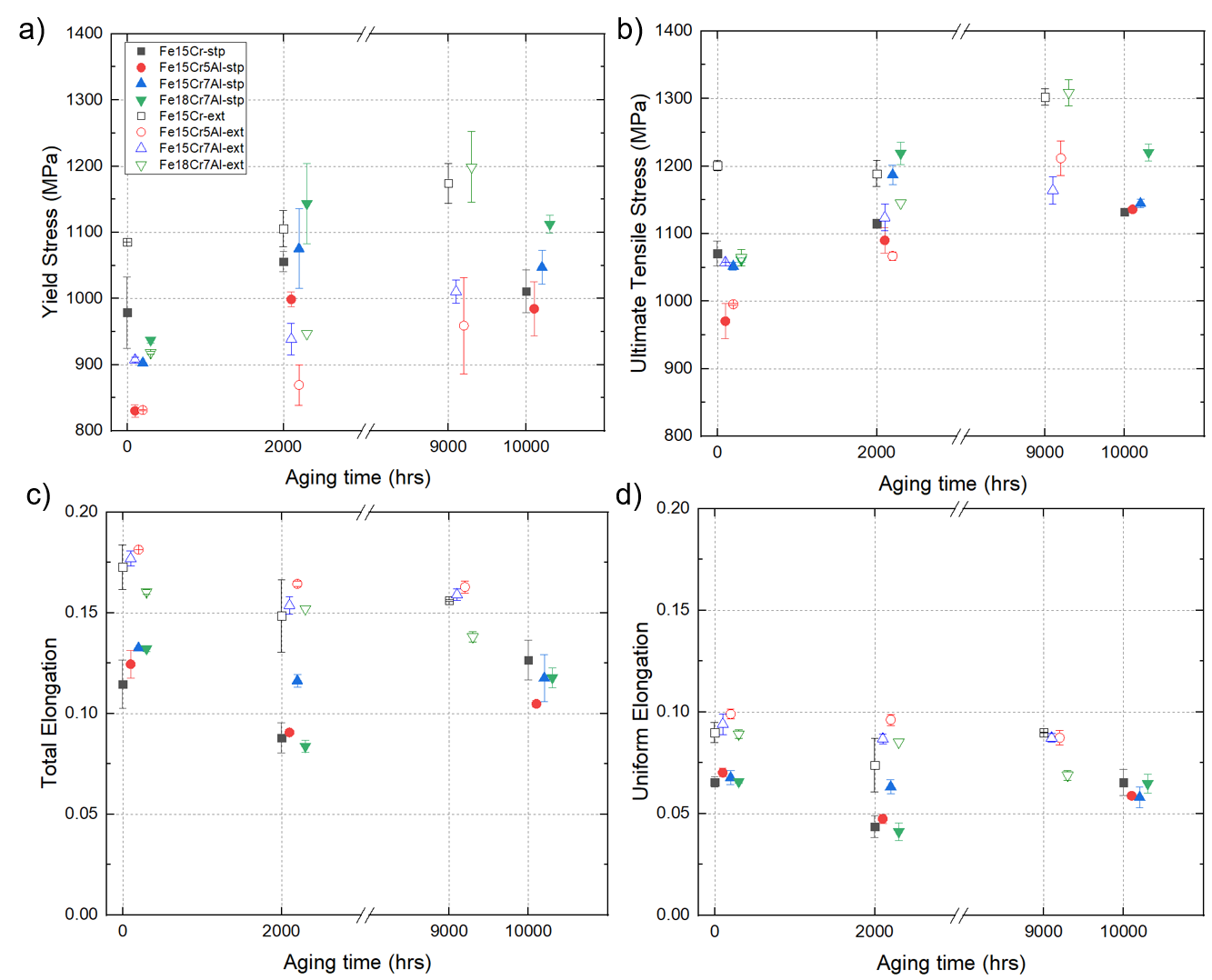}
\caption{The a) yield strength (YS), b) ultimate tensile strength (UTS), c) total elongation (TE), d) uniform elongation (UE) of ODS ferritic steels as-extruded (hollow) and after STP (solid), and after aging for both the specimens as extruded and after STP.}\label{fig:6}
\end{figure*}

\begin{figure*}[btp] 
\centering
\includegraphics[width=0.6\textwidth]{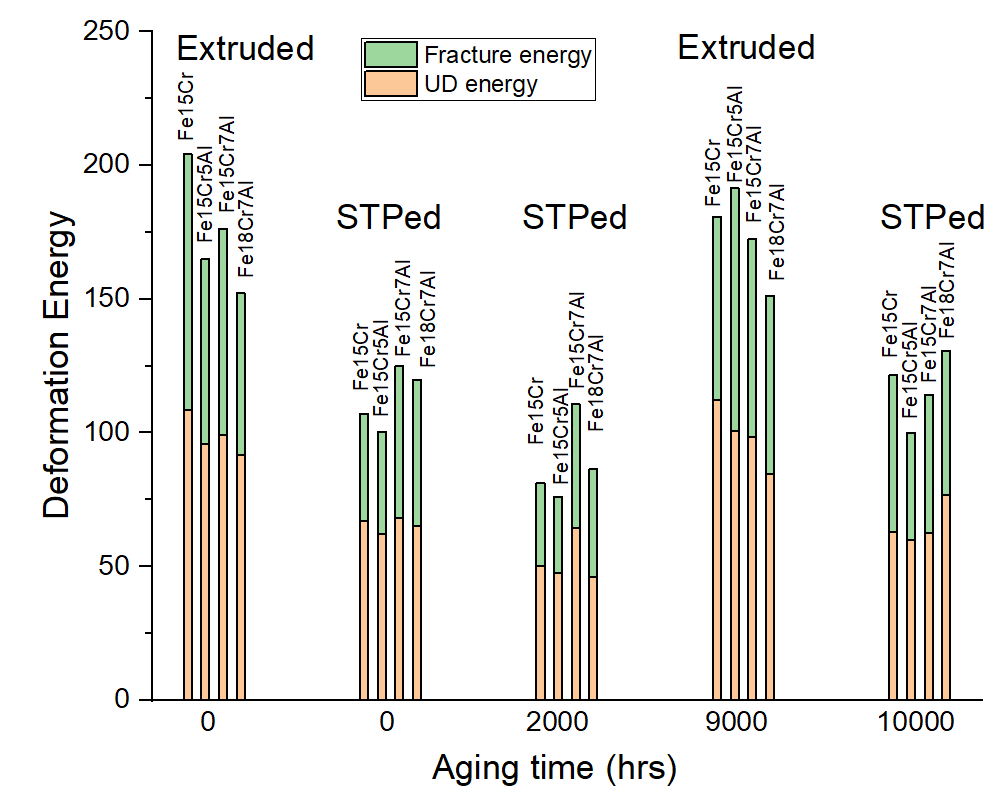}
\caption{Deformation energy ($J\cdot m^{-3}$) of as-extruded and STPed steels before and after $475\textdegree C$ aging.}\label{fig:7}
\end{figure*}

\begin{figure*}[btp] 
\centering
\includegraphics[width=0.9\textwidth]{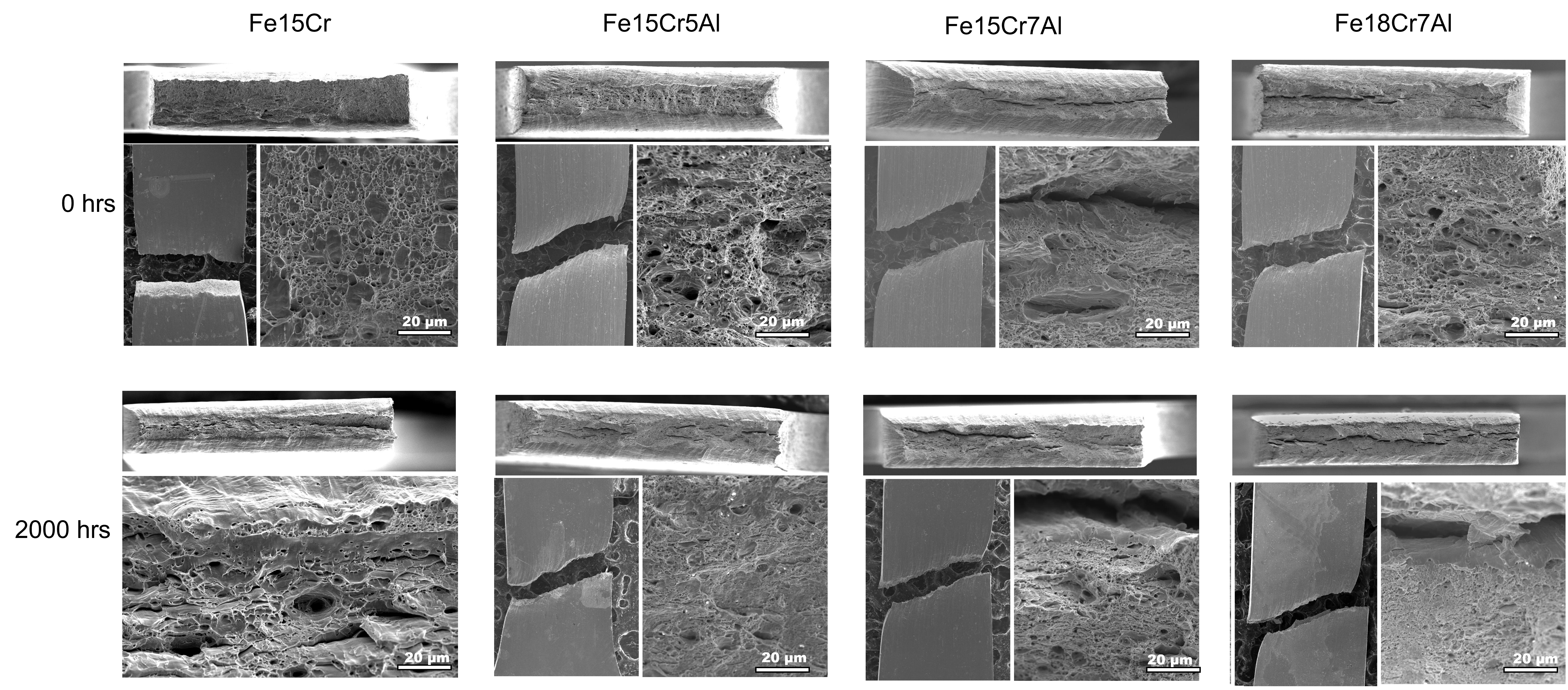}
\caption{The morphology of fracture surface of STPed Fe15Cr, Fe15Cr5Al, Fe15Cr7Al and Fe18Cr7Al before and after 2000hrs aging.}\label{fig:8}
\end{figure*}

\section{Experiments}\label{sec:2} 
\subsection{Materials}\label{subsec:2.1} 
The starting materials used in this study are FeCrAl ODS ferritic steels (SP series) produced by KOBELCO, Ltd. The chemical compositions are shown in Table \ref{tab:1}. There are four steels with different Al and Cr concentration, which are named as Fe15Cr (SP2), Fe15Cr5Al (SP4), Fe15Cr7Al (SP7) and Fe18Cr7Al (SP11). The Fe15Cr (SP2) steel contains no Al but 2wt\% W. All the steels were produced by mechanical alloying method starting with Ar-gas-atomized alloy powders, element powders and Y\textsubscript{2}O\textsubscript{3} powders. The mechanically alloyed powders were encapsuled into a can to degas at 400\textdegree C in a vacuum of 0.1 Pa for 2 hrs and extruded at 1150 \textdegree C into a rod with 25 mm diameter. The rod was finally annealed at 1150 \textdegree C for 1 hr and subjected to air cooling.

\begin{table*}[h]
\centering
 \caption{Chemical compositions of FeCrAl ODS ferritic steels (wt\%, Bal. Fe)}\label{tab:1}
 \begin{tabular}{l c c c c c c c c cc}
 \hline 
  & ID & Cr & Al & W & Ti & Y & C & O & N & Ar \\ 
 \hline 
 Fe15Cr & SP2 & 14.24 & $<$0.01 & 1.85 & 0.23 & 0.18 & 0.028 & 0.12 & 0.005 & 0.006 \\ 
 Fe15Cr5Al & SP4 & 14.39 & 4.65 & - & 0.32 & 0.39 & 0.032 & 0.22 & 0.005 & 0.006 \\ 
 Fe15Cr7Al & SP7 & 14.13 & 6.42 & - & 0.51 & 0.38 & 0.032 & 0.22 & 0.005 & 0.006 \\ 
 Fe18Cr7Al & SP11 & 16.83 & 6.31 & - & 0.49 & 0.38 & 0.032 & 0.22 & 0.004 & 0.006 \\ 
 \hline 
 \end{tabular}
\end{table*}

The workflow of the STP and aging experiments are illustrated in Figure \ref{fig:1}. The simulated tube fabrication process was started with the as-received ODS bars which were cut from the extruded rod. The bars were subjected to cold rolling and thermally annealing followed by furnace cooling at a rate of $\backsim$150 K\/hr in average to achieve a full ferrite phase. The CR-HT was repeated four times. The rolling process was illustrated in Figure \ref{fig:2}a. The rolling was performed at room temperature, with each single rolling pass yielding $\backsim$ 1 to 2\%  thickness reduction. The rolling direction (RD) was parallel to the extrusion direction without being reversed. As the specimens were very hard, cracks were easily generated at the head of the specimen during the rolling. In this case, we cut off these cracked parts to prevent it from growing deeper in following rolling. The thickness of specimens was approximately 50\% reduced in each rolling cycle, with subsequent annealing treatment at 950\textdegree C, 850\textdegree C, 1150\textdegree C and 1150\textdegree C for 1 hr, respectively. The total thickness reduction $\eta$ and strain $\epsilon$ are defined as: $\eta=1-h/h_0$ , $\eta=ln(h_0/h)$.  The initial sample thickness, h\textsubscript{0}, was 4 mm. The parameters of h, $\eta$ and $\epsilon$ in cold rolling and subsequent annealing temperatures were displayed in Table \ref{tab:2}. Tensile specimens of dog-bone shape were produced from STPed plates of 0.3 mm thickness. Isothermal aging was performed on the dog-bone tensile specimens sealed in vacuum capsules at 475\textdegree C for 2000 hrs and 10000 hrs followed by iced water quenching.  

\subsection{Microstructural observations}\label{subsec:2.2}
Transmission electron microscopy (TEM) was performed on JEOL 2200 field emission TEM to investigate the microstructures of grain and oxide morphology. The foils for TEM observation were thinned by focused ion beam (FIB, Hitachi FB2200) to 250 nm, followed by flash polishing in a mixture of 5\% HClO\textsubscript{4} and 95\% CH\textsubscript{3}OH at a voltage of 30 V and temperatures between -30 \textdegree C and -65 \textdegree C.

Electron Back Scatter Diffraction (EBSD) was applied to investigate the grain morphology via an EDAX detector equipped on Zeiss Ultra-55 field-emission scanning electron microscopy (FE-SEM) with acceleration voltage of 10 to 15eV. The overall grain morphology was scanned on an area of 150 $\mu m$ x 150 $\mu m$ with a step of 1 $\mu m$ via conventional EBSD to identify recrystallization occurrence. Since grains in ODS steels are generally sub-micron size with small misorientations, the Transmission Kikuchi Diffraction (TKD)\citep{kellerTransmissionKikuchiDiffraction2013} was performed as well with a step size of 50 nm to show the mixture of fine grains and recrystallized grains after partially recrystallization.

\subsection{Mechanical tests}\label{subsec:2.3}
Micro-Vickers hardness was tested by HMV-2T (Shimadzu Corp.) with 2 kg load and 10 sec holding time at room temperature. The hardness was measured on the rolling surface of the plates after each cycle of CR-HT.

Tensile tests were performed on INTESCO 205X tensile assembly with a load cell of 5 kN. The dog-born miniaturized tensile specimens were sampled from the STPed plates with the loading direction parallel to the rolling direction. The gage geometry of the tensile specimen was 5 mm in length, 1.2 mm in width and 0.3 mm in thickness. The tests were carried out at a displacement rate of 0.2 mm/min, resulting in an initial strain rate of 6.67$\times$10\textsuperscript{-4}/s. The yield stress (YS) was defined as 0.2\% off-set flow stress. Two or three specimens were tested at each aging condition. All the tensile tests were performed at room temperature.

\begin{table*}
\centering
 \caption{The parameters of simulated tubing process} \label{tab:2} 
\begin{tabular}{l r r r r}
\hline 
 & Thickness, h (mm) & Total thickness reduction, $\eta$ & Total strain, $\varepsilon$ & Annealing temperature, T (\textdegree C) \\ 
\hline 
Initial & 4 & 0 & 0 & - \\ 
CR-HT1 & 2 & 0.5 & 0.69 & 950 \\ 
CR-HT2 & 1 & 0.75 & 1.39 & 850 \\ 
CR-HT3 & 0.6 & 0.85 & 1.90 & 1150 \\ 
CR-HT4 & 0.3 & 0.925 & 2.59 & 1150 \\ 
\hline 
\end{tabular} 
\end{table*}

\section{Results}\label{sec:3}
\subsection{Simulated tubing processing}\label{subsec:3.1}
According to our previous research\citep{haEffectColdRolling2019}, the grain morphology of the FeCrAl ODS steels after hot extrusion have a strong $\alpha$-fiber texture (RD\/\/$<110>$) structure parallel to the extrusion direction, and have fine and isotropic grain shape on the cross-sectional surface. It was also shown that the fibers contained very fine sub-grains with small crystal misorientations. The hardness of the as-extruded steels are available in our previous publication\citep{hanEffectCrContents2016}. 

The evolution of grain morphology after each cold rolling and heat treatment (CR-HT) is shown in Fig. \ref{fig:2}(b).  After the first cycle (CR-HT1), Fe15Cr7Al (SP7) and Fe18Cr7Al (SP11) were subjected to recrystallization. It is obvious that Fe15Cr5Al (SP4) recrystallized after the 2nd cycle of CR-HT. In Fe15Cr (SP2), which is an Al-free ferritic steel, the recrystallization didn’t occur until the 3rd cycle of CR-HT. After the fourth cycle of CR-HT, there were still fine grains remaining, indicating that the grains were only partially recrystallized. Fig.\ref{fig:2}(c) is a cross-section image (normal to RD) of Fe15Cr aged for 2000 hrs. As the grain boundaries are considered stable at 475 \textdegree C, they can represent for the grain sizes after the STP. In Fig.\ref{fig:2}(c), both fine grains and coarse grains coexisted in the same specimen. The high thermal stability during STP of the grain structure in Fe15Cr (SP2) was interpreted in terms of fine oxide particle dispersion with a very high number density in Al-free ODS steel. These oxides could hinder the movement of grain boundaries, so that elevate the on-set recrystallization temperature. Fig.\ref{fig:3} summarizes the typical morphology of oxides in Al-free and Al-added ferritic ODS steels. In Fe15Cr7Al and Fe18Cr7Al, the oxides are Y-Al-O type with larger size but smaller density than Y-Ti-O oxides in Fe15Cr. These results correspond to the fact that adding Al will coarsen the size of oxides in ODS steels during fabrication\citep{oonoPrecipitationVariousOxides2019}. The growth of oxides during STP could be ignored due to short annealing time\citep{oonoGrowthOxideParticles2017}.

The hardness change by recrystallization reflects the change in grain sizes, ignoring the influence of textures. Cold rolling will induce subgrain boundaries and multiply dislocations, while thermal annealing will relieve the residual stress and trigger nucleation and grain growth if recrystallization occurred. The Vickers hardness after each CR-HT cycle of each ODS steel is shown in Fig.\ref{fig:4}. The Fe15Cr5Al and Fe15Cr7Al showed a similar behavior in Vickers hardness change. Fe15Cr showed a dramatic decrease after the CR-HT3. Combined with the grain morphology in Fig. \ref{fig:2}(b), Fig.\ref{fig:4} indicates that all the steels heated at 1150\textdegree C will endure significant recrystallization in HT3. Although the recrystallization was designed to soften material for further thinning, all the specimens in this work were finally hardened after the entire STP procedure compared to the as-extruded steels. 

The final hardening in CR-HT4 is because of the repeating recrystallization retarded by intermediate recrystallization in HT3. According to Leng et al\citep{lengEffectsTwostepCold2012}, repeating recrystallization requires higher on-set temperature and has higher hardness than first recrystallization at the same annealing temperature. Explanation was made by means of experiments of EBSD orientation density function (ODF) analysis\citep{aghamiriMicrostructureTextureEvolution2020,naritaCharacterizationRecrystallization12Cr2013,naritaDevelopmentTwostepSoftening2004,ukaiAlloyDesignCharacterization2023} and the theory built on nucleation driving force and grain boundary migration rate\citep{humphreysRecrystallizationRelatedAnnealing2012}. First, recrystallization will produce $\lbrace 111\rbrace$ and $\lbrace 110\rbrace$  textures on the rolling plane. Cold rolling on the recrystallized $\lbrace111\rbrace <112>$ crystals will produce strong $\lbrace 100\rbrace <110>$ texture, which contains extremely low strain energy that the driving force for recrystallization is very low, too. Second, the grain boundary migration rate depends on both the misorientation between the recrystallized nuclei and the matrix\citep{hayakawaNewModelGoss1997}, and the pinning force of fine oxides. For Al-free steels, the oxides are extremely small so that the pinning force controls the migration rate. For Al-added steels, the pinning force will be smaller. The nucleus of $\lbrace 111\rbrace <112>$ has a misorientation below 45\textdegree on rolling surface $\lbrace 112\rbrace <110>$ and $\lbrace 111\rbrace <110>$ \citep{aghamiriMicrostructureTextureEvolution2020} that they may grow faster during HT4.

In practice, cracks easily occurred in the steels during CR2 and CR3, particularly for Fe15Cr, which has extremely high hardness during CR3. There are no cracks generated in CR4 as recrystallization in HT3 greatly reduced the yielding strength. The final hardness after HT4 is slightly higher than the as-extruded, which means the steels are only partially recrystallized. Through the designed STP, the aim to increase the final strength than conventional fully recrystallized steels were achieved.

\subsection{Tensile properties}\label{subsec:3.2}
The engineering stress-strain curves of the STPed ODS steels were shown in Fig.\ref{fig:5} with a presentative test at each aging condition. The age-hardening showed dependence on the content concentration. Before aging, the tensile strength of FeCr ODS is reduced by Al-addition, which is due to the reduction of density of oxide particles. The yield strength increases with increasing Al and Cr concentration because of solid solution strengthening\citep{ukaiSolidsolutionStrengtheningCr2021}. As for the aging effects, age-hardening is much more significant in Al-added ODS steels (SP4, 7, 11) than that in Al-free steel (SP2). This behavior is similar to the as-extruded ODS steels in which (Al, Ti)-rich $\beta^\prime$-phases (as well as $\alpha^\prime$-phases) were induced by the aging\citep{sangEarlystageThermalAgeing2020,douEffectsContentsTi2020,douAgehardeningMechanisms15Cr2020}. 

Fig.\ref{fig:6} shows the yield stress (YS), ultimate tensile stress (UTS), uniform elongation (UE) and total elongation (TE) of all the tested specimens. Generally, the scattering of YS is larger than that of UTS because the ill-defined proof stress could be easily affected by the stiffness of tensile machine and/or load cell especially for thin specimens. The scattering of TE in Fig.\ref{fig:6}c was affected by the deformation after necking. A notable variation in $ds(e)/de$ occurred after the UTS, which means that the area of cross section of the specimen shrank rapidly after the necking occurrence. There is an inconsistency occurred in the yield stress and Vickers hardness of STPed SP2. In Fig.\ref{fig:4}, the Vickers hardness of STPed SP2 was similar to the as-received (as-extruded), however, in Fig.\ref{fig:6}a, the yield stress of STPed SP2 is much lower (~100MPa) than as-extruded steel. This might be owing to the anisotropy grain morphology in the steels after hot-extrusion and STP. As for the aging effect on the tensile strength and tensile elongation, all the 2000 hrs aged FeCrAl ODS steels showed YS hardening and TE reduction. The 10000 hrs aging, however, results in “recovery” effect with reduced hardening and increased elongation compared to the 2000 hrs aged ones.

The true stress-strain ($\sigma$-$\epsilon$) curves were estimated from the plastic deformation using the following equations: $\epsilon=ln(e_s^P+1)$,$\sigma(\epsilon)=s\cdot(e_s^P+1)$ where $e_s^P$ is the plastic component of engineering strain of specimen, $s$ is the engineering stress, $\epsilon$ and $\sigma(\epsilon)$ stands for the true strain and true stress respectively. The true stress-strain curve mainly deviates 1) after necking where the cross area reduced dramatically and the materials experienced three-dimensional stress state, and 2) around the initial plastic deformation regions where $e_s^P/e_s^E\ll1$. The empirical Ludwik relationship was evaluated by the data between 0.5\%  true strain to true uniform elongation:
\begin{equation}\label{eq:1}
\sigma(\epsilon)=\sigma_0+K\epsilon^n 
\end{equation}
where $\sigma_0$ is the true yield stress, $K$ is the strength coefficient, $n$ is the strain hardening exponent

The estimated $K$ and $n$ values from true stress-strain curves are listed in Table \ref{tab:3}. $K$ is the amplitude of the strain hardening term and $n$ is applied on the strain $\epsilon$ directly. Note that as the strain is smaller than 1, lower $n$ will lead to higher strain hardening ratio. From Table \ref{tab:3} we can conclude that the 2000 hrs aged specimens have the lowest $K$ and $n$ compared to other conditions. 

\begin{table*}
\centering
\caption{The strength coefficient $K$ and strain hardening opponent $n$ in Ludwik relationship}\label{tab:3}
\begin{tabular}{l c c c c c c}
\hline 
 & \multicolumn{2}{c}{0 hr} & \multicolumn{2}{c}{2000 hrs} & \multicolumn{2}{c}{10000 hrs} \\ 
 & $n$ & $K$ & $n$ & $K$ & $n$ & $K$ \\ 
\hline 
Fe15Cr & $0.355\pm0.024$ & $540.489\pm66.991$ & $0.313\pm0.043$ & $469.947\pm44.958$ & $0.305\pm0.019$ & $574.742\pm66.108$ \\ 
Fe15Cr5Al & $0.454\pm0.069$ & $872.585\pm78.248$ & $0.341\pm0.040$ & $559.785\pm33.790$ & $0.356\pm0.052$ & $749.861\pm60.764$ \\ 
Fe15Cr7Al & $0.376\pm0.010$ & $764.343\pm53.914$ & $0.345\pm0.053$ & $615.251\pm30.515$ & $0.418\pm0.034$ & $728.033\pm44.502$ \\ 
Fe18Cr7Al & $0.345\pm0.052$ & $674.347\pm27.706$ & $0.266\pm0.033$ & $434.258\pm27.903$ & $0.359\pm0.028$ & $663.538\pm65.669$ \\ 
\hline 
\end{tabular} 
\end{table*}

The aging embrittlement can be described by the reduction of total deformation energy, u\textsubscript{DE}, in tensile test by:
\begin{equation}\label{eq:2}
u_{DE}=\int_0^{UE}\sigma\,d\epsilon+\int_{UE}^{TE}\sigma\,d\epsilon
\end{equation}

The first term in equation (\ref{eq:2}) represents the energy applied by uniform deformation (UDE) until tensile stress reaches the UTS. The second term is defined as the fracture energy (FE) which is accompanied by necking. In practice, the integration was calculated by trapezoidal method. In Fig.\ref{fig:7}, the STPed steels exhibited significant decreases in $u_{DE}$ compared to extruded steels irrespective of Al concentration in each ODS steel. This reduction should be owing to the micro-crack generation during cold rolling. The recrystallized layered grains may accelerate the growth of cracks due to weak grain boundary cohesion. 

As for the effect of aging on the DE, three types of trends were showed: 1) for the Fe15Cr, the STPed steels showed reduction in DE after 2000hrs aging, but recovered at 10000 hrs aging, which was even higher than the non-aged one. This behavior is different to the as-extruded steels, whose DE at 9000 hrs aging was still lower than non-aged. 2) for the Fe15Cr7Al, the DE of both STPed and as-extruded steels decreases as aging time increases. 3) for the Fe15Cr5Al and Fe18Cr7Al, the DE of STPed steels reduced at 2000 hrs but increased at 10000 hrs aging, which are similar to the as-extruded ones. The three types of behavior of DE correspond to the three different modes of precipitates induced by aging: 1) only $\alpha^\prime$ in Fe15Cr and 2) only $\beta^\prime$ in Fe15Cr7Al were generated, but 3) both $\alpha^\prime$ and $\beta^\prime$ precipitates formed in Fe15Cr5Al and Fe18Cr7Al, which will be discussed in Section \ref{subsec:4.2}.

\subsection{Fractography}\label{subsec:3.3}
All the specimens before and after aging fractured in a ductile manner with plastic shearing induced dimples on the rupture surfaces. This phenomenon indicates that the materials still behave as ductile after 10000 hrs aging in tensile test at ambient temperature. This behavior is different to the recent research which showed a typical brittle fracture manner of Fe15Cr7Al ODS steel after 15000 hrs aging with tensile loaded in the tube hoop direction\citep{yanoEffectsThermalAging2021}.

\begin{figure*}[btp] 
\centering
\includegraphics[width=0.8\textwidth]{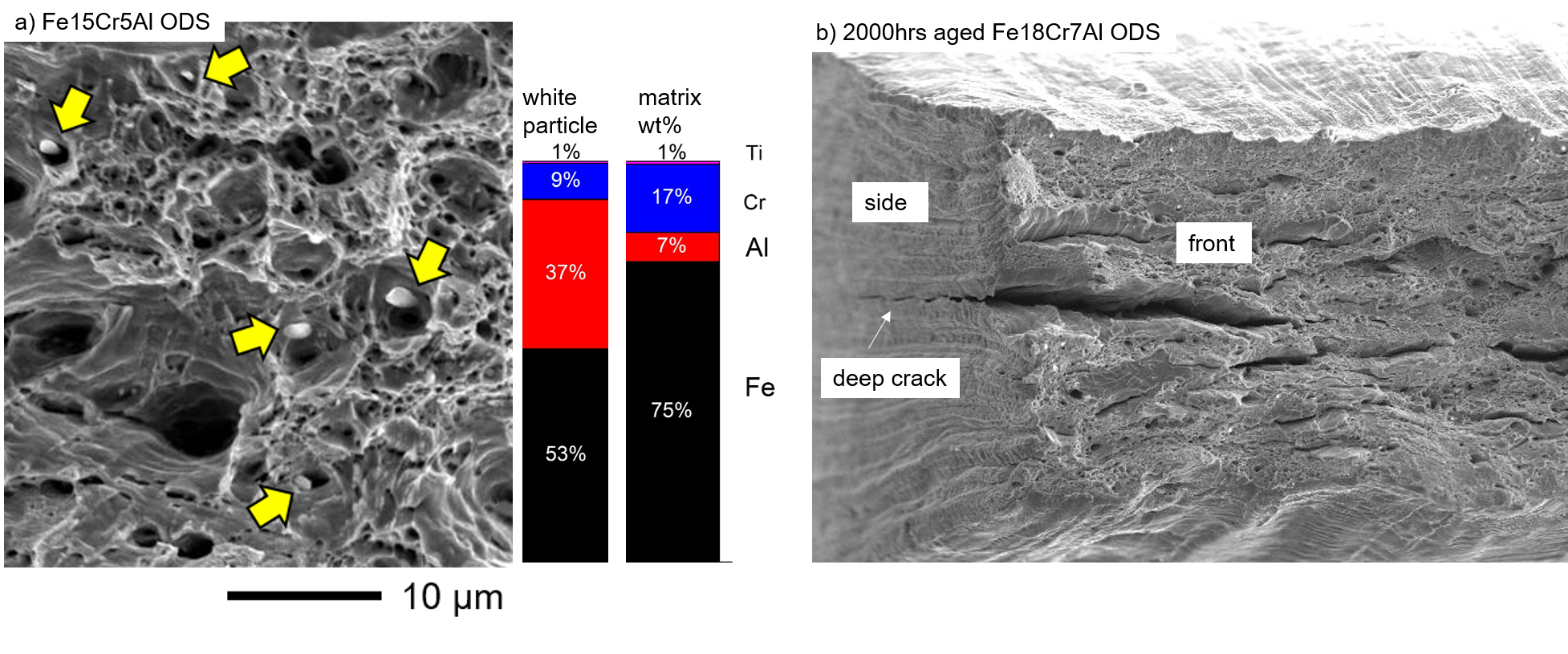}
\caption{The SEM images of fracture surfaces: a) Al-containing particles in dimples and b) secondary cracks penetrating along elongated grains.}\label{fig:9}
\end{figure*}

Fig.\ref{fig:9} shows two characteristic features on the fracture surfaces. The first one is large precipitates located inside the dimples as shown in Fig.\ref{fig:9}a. These precipitates are enriched in Al according to EDS spectrum analysis and can be deduced as alumina. These particles can function as the initial separation sites because of dislocation pile-ups and the weak adhesion of the interface between the particle and the matrix. The second one is the long secondary cracks as shown in Fig. \ref{fig:9}b. These cracks are formed on the layered and elongated grain boundaries, which are parallel to the rolling surface. This delamination is a typical phenomenon in laminated ODS steels\citep{kimuraDelaminationTougheningUltrafine2010}. The bamboo-like grains in the STPed steels will help the propagation of the secondary cracks\citep{dasWhySecondaryCracks2018}. The secondary cracks might be the reason for the elongation reduction compared to the as-extruded ODS steels.

\section{Discussion}\label{sec:4}
\subsection{The aging effect on total elongation}\label{subsec:4.1}
The fabrication of FeCrAl ODS cladding tubes requires recovery and recrystallization as a standard routine. Occurrence of recrystallization depends on the cold rolling degree and annealing temperatures which determines stored energy, namely, driving force of recrystallization. While Ha\citep{haRecrystallizationBehaviorOxide2014} reported that a full recrystallization would not cause sever loss of elongation in extruded ODS steels, the results from Yano et al\citep{yanoEffectsThermalAging2021} and Sakamoto et al\citep{sakamotoDevelopmentAccidentTolerant2021} showed a rather large ductility loss in pilger rolled recrystallized FeCrAl ODS steels. There is approximately 30\%  reduction of total elongation in steels after full tubing routine compared to as-extruded ones\citep{yanoEffectsThermalAging2021}. In this study, the STPed steels in Fig.\ref{fig:6} showed a similar trend as those after full tubing routine with a significant loss of elongation. Only a small loss of elongation after the recrystallization in the work by Ha\citep{haRecrystallizationBehaviorOxide2014} is probably due to no tubing process put on, suggesting that the loss of elongation is closely associated with cold rolling process. 

\begin{figure*}[btp] 
\centering
\includegraphics[width=0.7\textwidth]{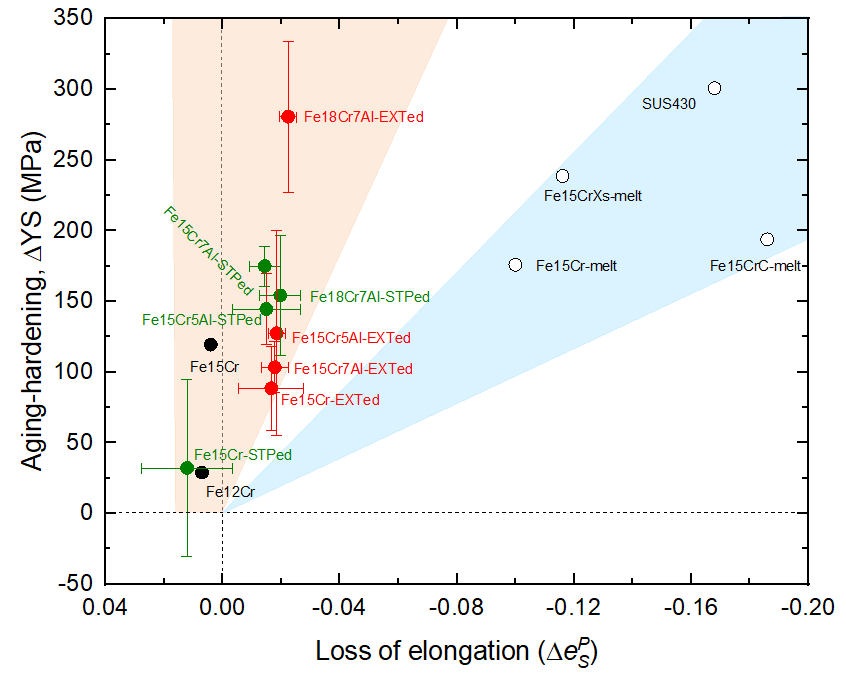}
\caption{Relationships between age-hardening and loss of elongation at the conditions of as-extruded (red), after STP (green), and arc-melted (non-ODS) alloys (hollow). The ageing temperature was 475 \textdegree C for all the specimens and the ageing period was 9000hrs and 10000hrs for as-extruded and STP, respectively. }\label{fig:10}
\end{figure*}

One of the supreme behaviors in the plastic deformation of FeCrAl ODS steels is their smaller loss of elongation by thermal aging compared to non-ODS steels\citep{chenIronChromiumPhase2015}. To illustrate this phenomenon, the aging-hardening in terms of the change in YS with respect to the loss of elongation of all the steels were shown in Fig.\ref{fig:10}. The Fe15Cr-melt, Fe15CrC-melt and Fe15CrXs-melt are non-ODS steels produced by arc-melting method. The Fe15Cr and Fe12Cr are ODS steels produced by mechanical alloying method\citep{chenCorrelationFeCr2014}. A commercial SUS430 which contains 16\% Cr without Al was compared as well. It should be noted that the aging effect were quite stable after approximately 5000 hrs aging according to the previous experiment results\citep{chenAgehardeningSusceptibilityHighCr2015}.

Fig. \ref{fig:10} shows that the loss of elongation by aging is rather smaller in ODS ferritic steels (red and green) than in arc-melted (non-ODS) alloys (hollow black) with respect to the same amount of age-hardening. This behavior is consistent with our previous works\citep{chenIronChromiumPhase2015,chenAgehardeningSusceptibilityHighCr2015}.  As for the comparison between as-extruded (red) and STPed (green) conditions, the loss of elongation in these steels are quite similar ($<$0.03). This result indicates that the ductility loss by aging is less sensitive to the nature of ODS steels (concentration, grain morphology, precipitates, etc).  

The rupture of ductile materials could be explained by micro-voids forming along the center of the necked region. The voids could form around precipitates where dislocation pile up and stress concentration occurs. Thus, non-deformable large particles, such as core-shelled Y-Ti-O in Al-free ODS steels, and $\beta^\prime$ and Y-Al-O in Al-added ODS steels, could act as initial nucleation sites of micro-voids. The theory to evaluate the ductile elongation developed by McClintock\citep{mcclintockCriterionDuctileFracture1968}:

\begin{equation}\label{eq:3}
\epsilon_{f}=\frac{\left(1-n \right)ln\left( l_{0}/2b_{0} \right)}{sinh\left[ \left( 1-n \right)\left( \sigma_{a}+\sigma_{b} \right)/\left( 2\overline{\sigma}/\sqrt{3} \right) \right]}
\end{equation}

Where $\epsilon_f$ is the strain to fracture, $n$ is the hardening exponent in Ludwik relationship, $l_0$ is the mean spacing of micro-voids, $b_0$ is the initial radius of holes, $\sigma_a$ and $\sigma_b$ are stresses related to different direction of holes, $\overline{\sigma}$ is the true flow stress.

It shows that the higher spacing of precipitates (micro-voids) and the higher strain-hardening exponent n in equation (1) will yield higher ductility. The latter is true that the lowest n at 2000 hrs aging in Table \ref{tab:3} also yields the lowest total elongation. The former is related to the density of various precipitates. As the total volume fraction of precipitates is insensitive to aging period at over-aging condition\citep{kimuraTwofoldAgehardeningMechanism2023}, density of precipitates will decrease as the radius increases, thus the fracture elongation will increase after the peak aging hardening.

The excellent resistance to aging ductility loss in ODS steels is owing to the smaller grain size in comparison to the non-ODS steels. As recrystallization occurs, the density of grain boundaries and triple junctions will be greatly reduced. These defects are effective dislocation sources at which the stress concentration will emit dislocations at initial plastic strain. The deformation at vicinity of grain boundaries is easier than grain interior due to higher number of mobile dislocations which might avoid the pile up at obstacles such as aging precipitates in center of grains. Therefore, the increased mobile dislocations emitted from triple junction and grain boundaries will contribute to the deformation of specimen during tensile thus showing enhanced resistance in elongation reduction by aging.

However, it should be also noted that grain boundaries could play a negative role in creep resistance at elevated temperatures. Under low stress that is less than the dislocation moving in center of grains, the deformation was mainly based on grain boundary sliding\citep{ukaiHightemperatureCreepDeformation2020}. Further, the emitted dislocations from triple junction could enhance the dislocation climb inside grains. It seems a trade-off existed in the grain size effect between aging resistance and creep resistance.

\subsection{The time dependent aging hardening}\label{subsec:4.2}
\begin{figure*}[btp] 
\centering
\includegraphics[width=0.7\textwidth]{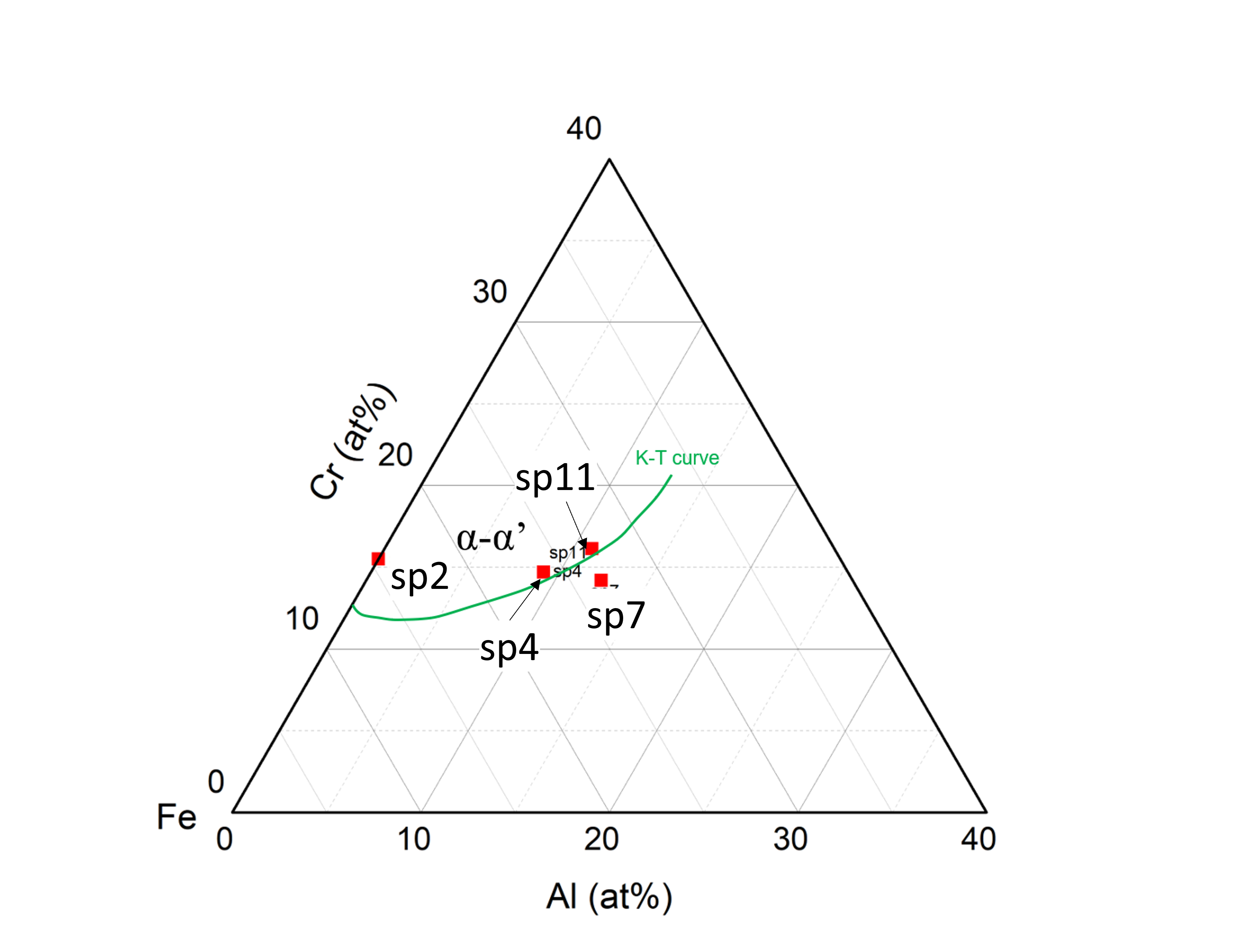}\label{fig:11}
\caption{The ternary diagram of Fe15Cr (SP2), Fe15Cr5Al (SP4), Fe15Cr7Al (SP7), Fe18Cr7Al (SP11). The Cr and Al concentrations are normalized atom ratios (Fe\% +Cr\% +Al\% =1). The superimposed green curve is the concentration boundary of $\alpha^\prime$ precipitation developed by Kobayashi and Takasugi (K-T).} 
\end{figure*}

\begin{figure*}[btp] 
\centering
\includegraphics[width=0.8\textwidth]{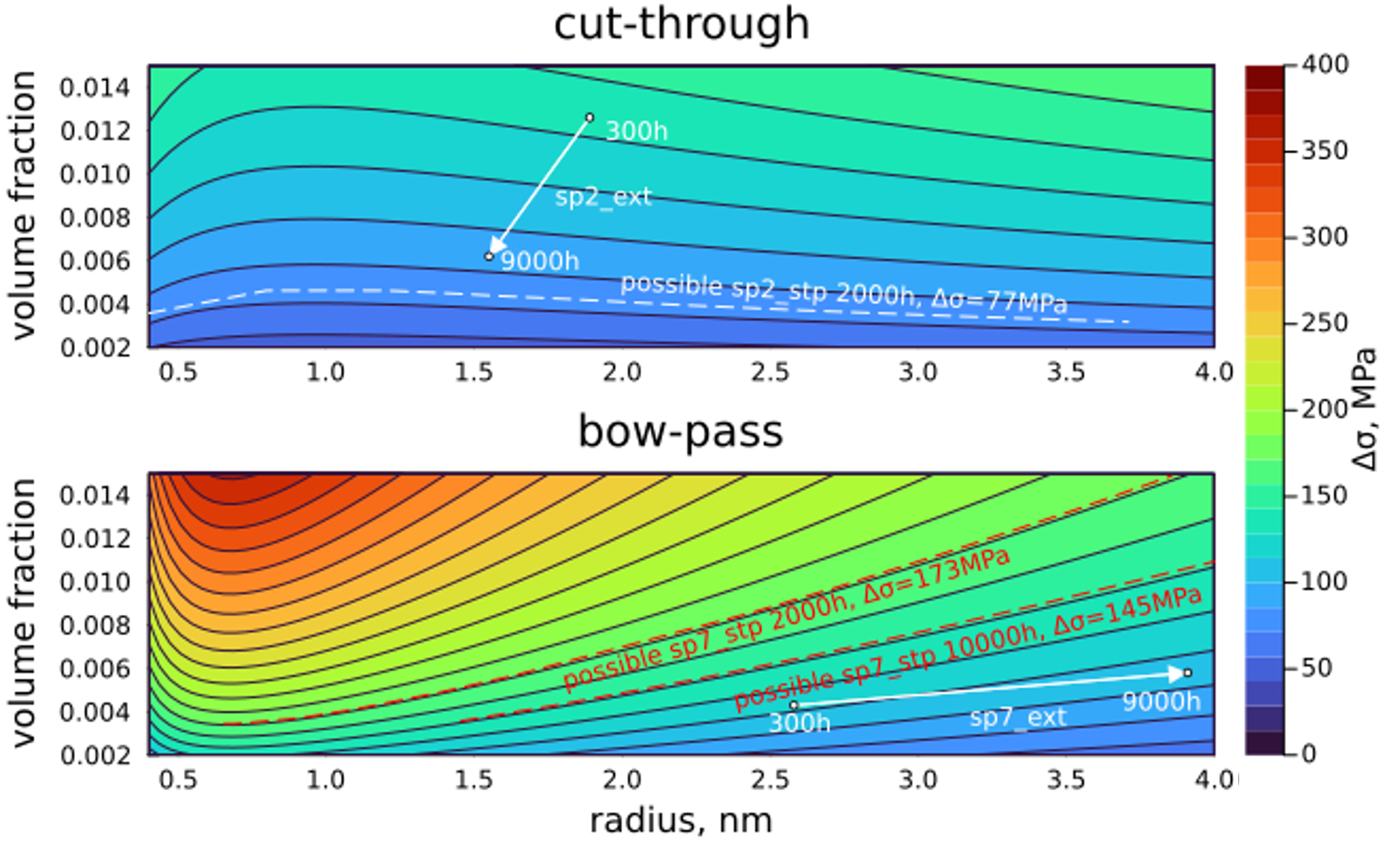}
\caption{The precipitate hardening by cut-through and bow-pass mechanism respectively. }\label{fig:12}
\end{figure*}

\begin{figure*}[btp] 
\centering
\includegraphics[width=0.6\textwidth]{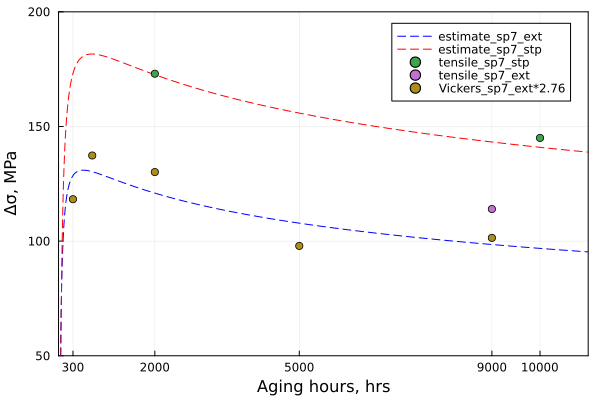}
\caption{The fitted aging hardening of Fe15Cr7Al (SP7).}\label{fig:13}
\end{figure*}
As aforementioned, the aging embrittlement was mainly ascribed to the formation of $\alpha^\prime$ and $\beta^\prime$ precipitates. Fig.\ref{fig:11} illustrated the $\alpha^\prime$ precipitation domain in the Fe-Cr-Al ternary phase diagram. The Fe15Cr (SP2) is in the inner region of $\alpha^\prime$ surrounded by the K-T curve. APT works have demonstrated the $\alpha^\prime$ precipitates formed since early stages of the aging\citep{sangEarlystageThermalAgeing2020}. The Fe15Cr5Al (SP4) and the Fe18Cr7Al (SP11) are located close to the K-T curve. The APT work showed the precipitates in Fe15Cr5Al contain both $\alpha^\prime$ and $\beta^\prime$ and core-shell structured oxides\citep{douEffectsContentsTi2020}. Currently there was no APT work for Fe18Cr7Al, but it could be speculated that the precipitates were similar to Fe15Cr5Al. The Fe15Cr7Al (SP7) is located off-shore of the $\alpha^\prime$ precipitation region. The APT work also demonstrated that no $\alpha^\prime$ but a number of $\beta^\prime$ precipitation formed after aging in this steel\citep{douEffectsContentsTi2020,douAgehardeningMechanisms15Cr2020}.

Dislocation interaction with precipitates are divided into two types 1) cutting-through and 2) bow-pass\citep{ardellPrecipitationHardening1985,martinPrecipitationHardeningTheory1998}. In the cutting-through mechanism, the coherent precipitates are deformable. The hardening of cutting-through could be expressed as:
\begin{equation}\label{eq:4}
\begin{cases}
&\Delta\sigma_{chs}=M\left( \frac{12}{\pi} \right)^{\frac{1}{2}}\gamma_{s}^{\frac{3}{2}}\left( \frac{f}{Gb} \right)^{\frac{1}{2}}\frac{1}{r}\\
&\Delta\sigma_{cohs}=M\alpha\left( \varepsilon G \right)^{\frac{3}{2}}\left( \frac{2rf}{Gb} \right)^{\frac{1}{2}}\\
&\Delta \sigma_{ms}=0.0055M\left( G_{p}-G \right)^{\frac{3}{2}}\left( \frac{2f}{G} \right)^{\frac{1}{2}}\left( \frac{r}{b} \right)^{0.275}\\
&\Delta\sigma_{cut-through}=\Delta\sigma_{chs}+\Delta\sigma_{cohs}+\Delta \sigma_{ms}\approx \Delta \sigma_{ms}
\end{cases}
\end{equation}

Where $\Delta\sigma_{chs}$ is the chemical strengthening related to the surface energy of precipitates, $\Delta\sigma_{cohs}$ is the coherency strengthening, $\Delta \sigma_{ms}$ is the modulus strengthening, $M=3.06$ is the Taylor factor, $\gamma_{s}$ is the interfacial energy between precipitates and matrix, $G\approx80$ GPa is the shear modulus of typical FeCrAl ODS steels, $G_p=115$ GPa is the modulus of $\alpha\prime$, $\varepsilon$ is the misfit of the interfaces.

The non-deformable hardening mechanism can be illustrated by the dispersoid barrier hardening model. A simplified Orowan type equation is used to describe this behavior:

\begin{equation}\label{eq:5}
\Delta \sigma_{box-pass}=0.1Gb\frac{f^{\frac{1}{2}}}{r}ln\frac{r}{b}
\end{equation}

Where $G$ is the shear modulus of matrix, $b$ is the Burgers vector of mobile dislocations, $f$ is the volume fraction of precipitates, $r$ is the radius of precipitates. 

A hypothesis was made in the following discussion that no supersaturation remained in the matrix, which means the volume fraction $f$ is close to a constant. This hypothesis is considered valid when aging over 2000 hrs.

When $f$ is constant and $r$ is larger than 1 nm, the derivative shows $\displaystyle \left.\frac{\partial \Delta\sigma_{cut-through}}{\partial r}\right|_f>0$, which means the aging hardening is monotone increasing. Similarly, $\displaystyle \left.\frac{\partial \Delta\sigma_{bow-pass}}{\partial r}\right|_f<0$, which means hardening by non-deformable precipitates will decrease with aging time (when $r>1nm$).

The critical radius of deformable $\alpha^\prime$ precipitates is around 4 nm, calculated by equations \ref{eq:4} and \ref{eq:5}. Above this critical radius, the cutting-through force becomes higher than bow-pass-by, that the $\alpha^\prime$ will become non-deformable. The actual critical radius may vary in different steels.

The hardening with precipitate radius and volume fraction by cut-through and bow-pass mechanism were shown in Fig.\ref{fig:12}. The hardening of cut-through mechanism is insensitive to particle radius. Therefore, the hardening in $\alpha^\prime$-only steels should be mainly induced by increasing of precipitates volume. This could explain the sever increase of hardening in as-extruded Fe12Cr and Fe15Cr in the first 2000 aging hours. After that, the radius of $\alpha^\prime$ gradually grew beyond the critical, the hardening mechanism converted to bow-pass. As the volume fraction remained nearly constant, the hardening decreased after the mechanism converting point. This could be supported by 5000 hrs aged as-extruded steels\citep{hanEffectCrContents2016} and the 10000hrs aged STPed Fe15Cr in this study, whose hardening are lower than 2000hrs aging, as shown in Fig.\ref{fig:6}a.

The hardening by bow-pass mechanism is sensitive to both precipitate radius and volume fraction. In the early stage of aging (~300hrs), $\beta^\prime$ started to appear with a rather large radius (2.58nm). The aging hardening went to maximum around 700 hrs, then started to decrease according to the Vickers hardness test\citep{hanEffectCrContents2016}. This indicates the volume fraction reached saturation in the coarsening (overaging) stage after 700hrs aging. The decreasing of hardening ascribes to the decreased number density of $\beta^\prime$ in Al-added ODS steels.

The coarsening process of precipitation may subject to the following kinetic equation\citep{ohayreMaterialsKineticsFundamentals2014}:

\begin{equation}\label{eq:6}
\left\langle R\left( t \right) \right\rangle^{m}-\left\langle R\left( 0 \right) \right\rangle^{m}=K_{r}t
\end{equation}

Where $\left\langle R\left( t \right) \right\rangle$ is the mean size of precipitates at time $t$, $\left\langle R\left( 0 \right) \right\rangle$ is the initial mean size, $K_r$ is the kinetic constant. In diffusion-limited coarsening, $m$ equals 3, and in source/sink-limited coarsening, $m$ equals 2\citep{balluffiKineticsMaterials2005}. 

In the analysis in PM2000\citep{capdevilaPhaseSeparationKinetics2012}, $\alpha\prime$ were found follows the LSW theory\citep{lifshitzKineticsPrecipitationSupersaturated1961,wagnerTheorieAlterungNiederschlagen1961}, where the radii of precipitates, $r$, could be expressed by a simplified temporal power law:

\begin{equation}\label{eq:7}
r=K_{r}t^{\frac{1}{3}}
\end{equation}

However, the kinetic constant $K_r$ and the exponent of time showed recrystallization dependence\citep{capdevilaPhaseSeparationKinetics2012}. Here we take the exponent of time as $1/3$ for both $\alpha^\prime$ and $\beta^\prime$. For simplicity, the change of oxides was ignored as well.

When only non-deformable precipitate exists, the combination of precipitation kinetics and hardening mechanism of equation \ref{eq:5} and \ref{eq:7} yields:

\begin{equation}\label{eq:8}
\Delta\sigma=0.1Gb\frac{f^{\frac{1}{2}}}{K_{r}}t^{-\frac{1}{3}}ln\frac{K_{r}t^{\frac{1}{3}}}{b}
\end{equation}

Hence the hardening is only dependent on aging time. Rewrite equation \ref{eq:8} we get:
\begin{equation}\label{eq:9}
\Delta\sigma\cdot t^{\frac{1}{3}}=a+klnt
\end{equation}

Where $a=0.1Gbf^{1/2}/K_{r}\cdot ln\left( K_{r}b \right)$ and $k=(1/30)Gbf^{1/2}/K_{r}$. The constant $K_r$ can be evaluated by $K_{r}=b\cdot exp(a/3k)$.

The fitting results from equation \ref{eq:9} were shown in Fig.\ref{fig:13}. The blue dash line is linear fitted by the Vickers hardness tests on aged bulk extruded materials, using the conversion equation $YS=2.76VH$. The STPed FeCrAl ODS steel has a higher hardening than as-extruded. This may be owing to the recrystallization which eliminated grain boundaries, where Ti segregates. Thus, the volume fraction of $\beta^\prime$ in STPed steel should be larger than in as-extruded, as predicted in Fig.\ref{fig:12}. 

The derivate of equation \ref{eq:9} with respect to $t$ yields:
\begin{equation}\label{eq:10}
\dot{\Delta\sigma}\cdot t^{\frac{1}{3}}+\Delta\sigma\cdot \frac{1}{3}t^{-\frac{1}{3}}=\frac{k}{t}
\end{equation}

Set $\dot{\Delta\sigma}=0$, combined with the equation \ref{eq:9}, the rest terms in equation \ref{eq:10} can be written as:
\begin{equation}\label{eq:11}
t_{max}=exp(3-a/k)
\end{equation}

This is the method to evaluate the over-aging time where maximum age hardening occurred for the non-deformable precipitates. The estimated kinetic constant $K_r$ and max hardening time $t_max$ are listed in Table \ref{tab:4}.

\begin{table}
\centering
\caption{The parameters $K_r$ and $t_max$ of aged FeCrAl ODS steels}\label{tab:4}
\begin{tabular}{ c c c }
\hline 
Fe15Cr7Al & $K_r (\times^{-3})$ & $t_{max}$(hrs) \\
\hline 
STPed & 7.385 & 691 \\ 
As-extruded & 5.515 & 513 \\ 
\hline 
\end{tabular} 
\end{table}

\section{Conclusion}
Four FeCrAl ferritic ODS steels, Fe15Cr (SP2), Fe15Cr5Al (SP4), Fe15Cr7Al (SP7) and Fe18Cr7Al (SP11), were fabricated by simulated tube processing (STP) to plates with 0.3 mm thickness. The plates were aged at 475\textdegree C for 2000 hrs and 10000 hrs in vacuum. Uniaxial tensile tests were performed to investigate the aging embrittlement in different steels. The obtained results are summarized as below:

\begin{enumerate}
	\item The four cycles of CR-HT STP with recrystallization in HT3 and repeating temperature in HT4 yielded partially recrystallization in the last STP step. 
	\item Tensile tests to the rolling direction showed that yield stress returned similar to the as-extruded ones after STP, except SP2 whose YS reduced 100MPa. The TE reduction of STPed steels is ~1/3 of the extruded steels. All the STPed steels showed reduction of deformation energy in tensile tests compared to extruded steels. 
	\item The deformation energy change after aging could be divided into three types, corresponding to the precipitation types of $\alpha^\prime$ and $\beta^\prime$ precipitates formation.
	\item For Al-free ODS steels after STP, the aging hardening is smaller than as-extruded ones. For Al-added ODS steels after STP, the aging hardening haviour is similar to as-extruded. Recovered effect in hardening and elongation reduction appeared after 10000 hrs aging.
	\item All the specimens/materials after aging fractured in a ductile manner. There were two characteristic features on the tensile fracture surface: 1) Al-containing particles were observed in the dimples and 2) secondary cracking along elongated layered grain.
	\item The loss of elongation by ageing is rather smaller in ODS ferritic steels than in non-ODS alloys with respect to the same amount of age-hardening. 
	\item Over-aging occurred before 10000 hrs annealing in all the STPed steels. The analysis combining Orowan hardening equation and LSW theory showed that the STPed SP7 has a higher $\beta^\prime$ growth rate constant.
\end{enumerate}

\section{Acknowledgement}
ZXZ would like to thank Prof. Akihiko Kimura in Kyoto University for the supporting of the experiment and the research communications.  




  \bibliographystyle{elsarticle-num-names} 
  \bibliography{My_Library.bib}
  \biboptions{sort&compress}

%
%
%
\end{document}